\documentclass[rmp,11pt,a4paper,showpacs,showkeys,preprint]{revtex4}
\usepackage{amsmath,amssymb,amsfonts}
\usepackage{graphicx}

\newcommand{\proof}{\par\medskip\noindent{\textit{Proof}:\,\,}}
\newcommand{\Rset}{\mathbb{R}}

\newcommand{\Pset}{\mathcal{P}}
\newcommand{\A}{\mathcal{A}}

\newcommand{\m}{\mathcal{M}}

\newcommand{\C}{\mathcal{C}}
\newcommand{\TL}{\mathcal{L}}

\newcommand{\tOm}{\widetilde{\Omega}}

\newcommand{\wf}{\widehat{F}}
\newcommand{\tp}{{\tilde p}}
\newcommand{\hf}{{\hat f}}

\newcommand{\tx}{{\tilde x}}

\newcommand{\txi}{{\tilde\xi}}
\newcommand{\tzeta}{{\tilde\zeta}}

\newcommand{\tmu}{{\tilde\mu}}

\newcommand{\ttau}{{\tilde\tau}}

\newcommand{\lvec}[1]{{\stackrel{\longrightarrow}{#1}}}

\newtheorem{definition}{Definition}

\begin{document}
\baselineskip=16pt

\title{Conformal proper times according to the Woodhouse causal axiomatics of relativistic spacetimes}

\author{Jacques L. RUBIN}
\email{jacques.rubin@inln.cnrs.fr}

\affiliation{
Universit{\'e} de Nice -- Sophia{-}Antipolis, UFR Sciences,\\
Institut du Non-Lin{\'e}aire de Nice -- CNRS, UMR6618,\\
1361 route des Lucioles, 06560 Valbonne, France}

\begin{abstract}
On the basis of the Woodhouse causal axiomatics, we show that conformal proper times and an extra variable in addition to those of space and time, precisely and physically identified from experimental examples, together give a physical justification for the `\textit{chronometric hypothesis}' of general relativity. Indeed, we show that, with a lack of these latter two ingredients, no clock paradox solution exists in which the clock and message functions are solely at the origin of the asymmetry.
These proper times originate from a given conformal structure of the spacetime when ascribing different compatible projective structures to each Woodhouse particle, and then, each defines a specific Weylian sheaf structure. In addition, the proper time parameterizations, as two point functions, cannot be defined irrespective of the processes in the relative changes of physical characteristics. These processes are included via path-dependent conformal scale factors, which act like sockets for any kind of physical interaction and also represent the values of the variable associated with the extra dimension. As such, the differential aging differs far beyond the first and second clock effects in Weyl geometries, with the latter finally appearing to not be suitable.
\end{abstract}

\preprint{Preprint INLN \#2008/06/01 -- June the 25th -- 2008}

\date{\today}

\maketitle


\section{Clocks and freely falling particles in Woodhouse causal axiomatics}
Justifying its use in what follows,
one of the great advantages of the Woodhouse axiomatics  \citeyearpar{wood73} resides precisely in the absence of the chronometric hypothesis (CH) of general relativity as well as in references to somehow unspecified standard or ``absolute" clocks.
Let us recall that the chronometric hypothesis is defined by the equality $ds\equiv c\,d\tau$, where $\tau$ is the proper time of the clock (also known as the `\textit{atomic time}'), $ds$ is the time interval between two events and is defined from a given metric field $g$, and $s$ is the so-called `\textit{gravitational time}' (up to the speed of light $c$).
Ehlers \textit{et al.} \citeyearpar{ehl72} strongly criticized this hypothesis, invoking at least two reasons: 
1) it can be replaced by the conformal or projective structures without a metric field, and 2) the shifts between atomic and gravitational times can be deduced from the Kundt-Hoffmann protocol \citeyearpar{kundhoff62} and, as a result, allow the selection of those clocks satisfying the equality defined by the CH.
\par\smallskip
These axiomatics are one of those appropriate frameworks under which the latter shifts, which are scale factors differing from $c$, enable us to track physical justifications to the CH. The main tool we use throughout is the geometrical situation encountered in the so-called `\textit{clock (pseudo-)paradox},' the solutions of which explicitly require the CH. Additionally, this is another great advantage of the Woodhouse axiomatics for approaching  temporal paradoxes, since it does not refer to any proper time notions. 
\par\smallskip
For more on the geometric settings, we recall that in these axiomatics, the so-called `\textit{message functions}' $f_p^\pm$  and  `\textit{clock functions}' $t_p$ (i.e. the time parameterizations defined by Woodhouse) are only defined for `\textit{particles}' $p$ ascribed to worldlines of freely falling massive punctual objects, and the metric fields $g$ depend only on particles $p$ up to conformal factors defined for clock functions $t_p$.
In addition, upstream of the metric structure given by $g$, there is a unique affine structure given by the totality of the freely falling particles of the Woodhouse axiomatics or equivalently, from Ehlers \textit{et al.}, a Weyl structure, i.e., in particular, a class of conformally equivalent Riemann structures. Weyl structures do not fix rates of clocks a priori independently of their histories. This is the so-called second clock effect. The opposite case was dismissed by Audretsch \citeyearpar{audret83}, who concluded that there were several efforts after 1970 to assign a Weyl geometry to spacetime rather than a Riemannian geometry. In reality, this is close to our viewpoint, i.e., that rates of clocks, invoked to approach the CH via the clock paradox, are not observable, but fixed only respective to very particular given causal protocols exhibiting the physical and historical relationships among particles themselves, each endowed with particular different sub-structures of a given common conformal one. Obviously, whenever the CH is posed and the metric connection is chosen, then the clocks with proper time aging (or those that are ``displaying" a proper time) are automatically inferred from the so-called `geodesic hypotheses' as well as singled out, and so, there are clearly no clock paradoxes.
The differential aging is quite simply given from the proper time differences obtained from the integration of the infinitesimal proper times along each worldline. However, the CH actually raises the point that we must feature a proper time notion in the physical viewpoint. That is, we must find a proper time definition \textit{regardless of} the choice of time parameterizations of the worldlines carried out for each particle of the Woodhouse axiomatics. On the contrary, the differential aging would only translate or reflect the agreement between the proper time definition, and, in particular, the CH and the selected metric connections.
That is to say, this agreement expresses the metric structure choice from the unique affine structure of a Weyl spacetime.
\par\medskip
Additionally, the Woodhouse results are situated upstream of the projective, conformal and Weyl structures defined in particular in the Ehlers \textit{et al.} axiomatics. Terms such as `\textit{freely falling particles}' are simply generic, all the more so as their worldlines cannot be defined from geodesic equations since metric connections are never utilized in the Woodhouse formalism. 
The only important restriction in the definition of the particles is mainly, in our opinion, that it must satisfy the following fundamental property: that for each particle $p$ passing through each event $e$ in spacetime, there exists a neighborhood of $e$ into which a four-dimensional `\textit{$C^0$-congruence}' can be defined and which can also be an extension of $p$. To some extent, this condition is a topological translation of the possible existence in this neighborhood of a projective connection. It can finally be substituted for Axiom $\mathrm{P}_2$ of Ehlers \textit{et al.}, specifying the (projective) equations assigned to freely falling particles and, in some ways, incidentally and unfortunately, making a contradictory use of proper times. This indicates another advantage of the Woodhouse axiomatics.
As a fundamental result, the true freely falling particles are those belonging to a \text{$C^0$-congruence}. Particles subjected to forces do not belong to it, but even in this case, the message and clock functions are still usable in the Woodhouse formalism.
\par\medskip
Returning to the tool used in the following, a resolution of the clock paradox is possible if the accelerations are involved in the differential aging formulas, since they are the sole data distinguishing the two clocks.
This accounts for the well-known historical attempts made by Einstein \citeyearpar{Einstein18} to solve it in general relativity. He used, in his own terms, a `\textit{pseudo-gravitational}' field accounting for the accelerating force experienced by the moving clock, which is at the origin of the clock asymmetry. Unfortunately, we run into an important problem with at least two distinct physical situations: the first is met in the case of rectilinear motions and the second is related to circular motions. In the first case, we refer, for example, to the Unnikrishnan \citeyearpar{unnik2005} results, recalling and clearly showing some possible inconsistencies in the solutions based upon general relativity due to the use of pseudo-gravitational fields. 
However, in the second situation for circular motions and among the most precise recent experimental results, differential aging clearly does not depend in any way on a pseudo-gravitational field such as, for instance, the pseudo-field associated with the centrifugal force. In this last case, the conclusion is clear: special relativity is more than sufficient to fully account for the experimental results.
Therefore, it is necessary to raise the question of why, when motions are rectilinear, we must use general relativity, if that is indeed what must be done, and why, in the case of circular motions, general relativity no longer intervenes. Nevertheless, a certain type of mechanical work might also be used for interpretations and solutions, as we shall see. It is a fact that in the first, linear case, some forces work, whereas in the case of circular motions, centripetal forces do not.
\par\medskip
In addition, at the origin of the conformal proper times, considering an accelerating frame under a kind of primary initial geometry or special relativity, we know, for instance, that there are precession effects influencing space vectors, such as the spin vectors leading to the so-called Thomas precession.
The latter has a long-standing history in relation to the clock paradox. At the basis of the precession equations, there is, as a particular case, the so-called \textit{M{\o}ller term} $u\wedge\dot{u}$, where $u$ is a velocity vector, or more generally, if $u\equiv U$ is a velocity field, \(U\wedge\nabla_UU\). Then, if $u$ is the restriction of $U$ to a worldline and is tangent to the latter, this M{\o}ller term can exactly account for another metric connection restricted to this worldline. As a result, the precessing vector is the tangent vector of a geodesic. Furthermore, if $U$ precesses itself with respect to another precession equation, i.e., if $U$ is \textit{congruent} with another vector field, then its  dual with respect to the metric defines a \textit{Weyl proper time} \citep[Remark (d) p.67, and p.81]{ehl72}. Moreover, the vanishing behavior of this M{\o}ller term --meaning the primary connection $\nabla$ is a projective connection with its corresponding geodesics-- defines the projective geodesics as being those of the primary initial geometry. In other words, a non-vanishing M{\o}ller term points to a change in a non-equivalent projective geometry, i.e., a change of physical frame of reference, but it is at least compatible with the conformal structure. Consequently, again and in particular, the flat Minkowski spacetime should no longer be valid for non-geodesic motions.
We could say that the projective connections and, thus, the Weyl structures depend on the non-geodesic motions to which they apply. This is a way of justifying the use of varying Weyl structures (or projective structures) attached to each particle. Then, each worldline or particle should correspond to the restriction of a specific Weyl structure only, and all these non-projectively equivalent Weyl structures should be conformally related, leading to (relative) conformal proper times.
\par\smallskip
In the next section, we recall the different definitions given in the Woodhouse axiomatics and the Malament Theorem. Section \ref{coeur} is divided into four sub-sections, of which the second is devoted to a difficulty that occurs systematically and is inherited from the simultaneity maps resulting in a non-trivial differential aging. In the third sub-section, geometrical arguments that are necessary to define formulas for differential aging and are consistent with the experimental results are presented using conformal proper times. In the fourth sub-section, the conformal factors in the previous formulas are physically interpreted, and then in section \ref{expert}, these formulas are applied to certain experimental results.
\section{Woodhouse causal axiomatics and the Malament
Theorem}
In the Woodhouse axiomatics, the spacetime $\m$ is a set of points, to which a set $\Pset$ of subsets of $\m$ called `\textit{particles}' and denoted by $p$ is associated, each of them being homeomorphic to $\Rset$. Thus, the set $\m$ is not a priori a topological manifold, and the particles cannot be loops. It is assumed that at least one particle $p$ passes through each point $x$ of $\m$. The first axiom of causality, i.e., Axiom 1a, states that, for each particle, an orientation can be selected among the two determined by the homeomorphism with $\Rset$ such that, once these orientations are selected and fixed for the whole set $\Pset$ of the particles, each event $x$ of $\m$ cannot be chronological to itself, i.e., $\forall\,x\in\m$ then $x\not\!\ll x$. The `\textit{chronological relation $\ll$\,}' is then a partial and anti-reflexive order on $\m$. This is a global property of $\m$ called the
`\textit{Chronological Principle}', i.e., there are no causal loops, and thus, $\m$ is also called a `\textit{chronological}' spacetime.
Axiom 2 states that the intersection of any particle $p$ with any open $I^\pm(x)$ of the Alexandrov topology\footnote{In the Woodhouse axiomatics, one recalls that if $x\in\m$ then $I^+(x)=\{y\in\m\,/\,x\ll y\}$\,, and also $I^-(x)=\{y\in\m\,/\,y\ll x\}$. These sets are open for the Alexandrov topology.}  is open for the topology on $p$ (issued from the Borel topology on $\Rset$). From Axioms 1a, 1b, 2 and 3, providing $\m$ with the Alexandrov topology $\A$, then $(\m,\A)$ is Hausdorff, the chronological relation $\ll$ is 
 `\textit{past- and future-distinguishing'}\,\footnote{The relation $\ll$ is `\textit{future distinguishing}' if $I^+(x)=I^+(y)\Longrightarrow x=y$, and `\textit{past distinguishing}' if $I^-(x)=I^-(y)\Longrightarrow x=y$.}
and `\textit{full'},\,\footnote{That is: 1) $\forall\,x\in\m$, $\exists\,y\in\m$ such that $y\ll x$, 2) if $y_1\ll x$ and $y_2\ll x$ then $\exists\,z\in\m$ such that $z\ll\,x$,\, $y_1\ll\,z$ and $y_2\ll\,z$, and 3) the dual relations of 1) and 2) are satisfied with the dual chronological relation $\gg$.}
and each event $x\in\m$ possesses a `\textit{past and future reflecting'}\,\footnote{An open neighborhood (for the Alexandrov topology $\A$) $N_x$ of $x$ is an `\textit{future and past reflecting}' open if $\forall\,y,\,z\in N_x$ then 1) $I^+(y)\supset I^+(z)\Longrightarrow I^-(z)\supset I^-(y)$ (\textit{future reflecting}), and 2) $I^-(y)\supset I^-(z)\Longrightarrow I^+(z)\supset I^+(y)$ (\textit{past reflecting}).
} open (for $\A$) neighborhood.
\par\medskip
Then, Woodhouse shows that $\m$ is of the Kronheimer-Penrose type
\citeyearpar{kronpen67}, where $\,<\,$ is the 
`\textit{causal relation}'\,\footnote{$x<y$ if $I^+(x)\supset I^+(y)$ and $I^-(x)\subset I^-(y)$.}
and $\uparrow$ the 
`\textit{horismotic relation}'\,\footnote{$x\uparrow y$ if $x<y$ and $x\not\!\ll y$.}
or `\textit{horismos}.'
Moreover, the chronological relation is transitive, which is not initially required in the Kronheimer-Penrose causal axiomatics in full generality.
\par\smallskip
The `\textit{message functions'\,\footnote{$\forall z\in\m$, $f^+_p(z)=\inf_{x\in p}\{x\,/\,z\ll x\}$ and $f^-_p(z)=\sup_{x\in p}\{x\,/\,x\ll z\}$.}\,} $f^\pm_p:U_p\longrightarrow p$ associated with each particle $p$ are defined on the tubular past- and future-reflecting Alexandrov open neighborhoods $U_p$, such that
\begin{equation}
U_p\equiv\bigcup_{m_1,m_2\in p}\!\!\!{<}m_1,m_2{>}, 
\end{equation}
where  ${<}m_1,m_2{>}=I^+(m_1)\cap I^-(m_2)$.
These functions are unique and increase strictly monotonically for the chronological relation, open and continuous on $U_p$.
\par
The `\textit{clock functions}' associated with each particle $p$ of $\Pset$ are homeomorphisms $t_p:p\longrightarrow\Rset$, which increase monotonically for the chronological relation. Additionally, we define the `\textit{radar coordinates}' $r^\pm_p\equiv t_p\circ f^\pm_p$ defined for 
each particle $p$. From this point on, for each pair of non-coplanar particles $p_1$ and $p_2$  (if coplanarity has a meaning in $\m$), Axiom 4a is satisfied if $\forall z\in\m$, $\exists\, p_1,\,p_2$ such that $z\in U_{p_1}\cap U_{p_2}$, and the four radar coordinates $r_{p_i}^\pm$ ($i=1,2$) together define a morphism:
\begin{equation}
U_{p_1}\cap U_{p_2}/{p_1\cup p_2}\longrightarrow U\,\mbox{open}\subseteq\Rset^4,
\end{equation}
which is a one-to-one map. Then, from this Axiom, it can be proven that $\m$ is a topological manifold homeomorphic to $\Rset^4$.
\par
In addition, from the `\textit{$C^0$-congruence}'\,\footnote{A set $\C=\{(p_\lambda,t_\lambda)\,/\,\lambda\in\Lambda\}$ of particles $p_\lambda$ with the parameterizations $t_\lambda:p_\lambda\longrightarrow\Rset$, is a \textit{$C^0$-congruence} on $U\subset\m$ if: 1) $\forall x\in U$, $\exists!(p_x,t_x)\in\C$ such that $x\in p_x$ and $t_x(x)=0$, and 2) $\forall\,V\subset\Rset$ a neighborhood of $0$, then $\forall r\in V$  the `\textit{evaluation map}' $E_r:x\in U\longrightarrow t_x^{-1}(r)\in\m$ is a homeomorphism.}
on a set $U\subseteq\m$ and from the\linebreak $C^1$-differen\-tia\-bi\-li\-ty assumption\footnote{Under conditions in which $U_p$ have no caustics, that is each message function $f^\pm_q$ restricted to $p$ with $q\subset U_p$ are diffeomorphisms.} of the clock and message functions, Woodhouse defined $C^n$-congruences and inductively proved that $\m$ can be assumed to be \textit{smooth}, i.e., of class $C^\infty$.
\par\smallskip
Lastly, we call `\textit{scalar potentials of metric $g_p$}' smooth functions defined on $U_p$ with values in $\Rset$, each one associated with a particle $p$, such as $\forall\,z\in U_p$:
\begin{equation}
g_p(z)\equiv r^+_p(z)\, r^-_p(z).
\end{equation}
Then, the Lorentzian metric $g$ on $\m$ is the Hessian of $g_p$: $g\equiv H(g_p)$ which no longer depends, nonetheless up to a conformal factor, on $p$ and $t_p$.
\par\smallskip
In addition, if $y$ is in a past and future reflecting open set of $x$ and $x\uparrow y$, then there exists \citep[Lemma 4.2]{wood73} a neighborhood $N_y$ of $y$ such that $N_y\cap\partial\bar{I}^{+}(x)\cap\partial\bar{I}^{-}(y)$  (with $\bar{I}^\pm=\{y\in\m\,/\,I^\pm(y)\subset I^\pm(x)\}$) is a (unique) light path and a $C^\infty$ one-dimensional submanifold. Each light path is a null curve, and they are null conformal geodesics of class $C^\infty$. Hence, if $x\uparrow y$ then there exists a unique (no caustics) light path that is at least piecewise smooth. 
\par 
One can notice that in the Woodhouse axiomatics, we may have $x\uparrow y$ without light path joining $x$ and $y$, meaning the latter have no common past- and future-reflecting open neighborhoods.
Then, defining a new horismos $\to$ such that $x\to y$ if $x\uparrow y$
and $x\not\!\ll y$, this new causal structure on $\m$ (or each $U_p$) is somehow ``natural", i.e., $\m$ is a conformal manifold with its field of light cones, such that $\to$ relates events on a cone or equivalently by a light path. And then, $\uparrow$ is identified with $\to$\,.
\par\medskip
Based on this topology, the projective, conformal and Weyl structures can be defined from the choices of metric connections, together with the constraints as indicated in the previous section. That is to say, the particles must be geodesic for the Weyl structure.
Moreover, we recall the Malament Theorem \citeyearpar{malam77}, which can be used further with the topology provided by the Woodhouse causal axiomatics:
\begin{quotation}
\noindent Malament Theorem -- \textit{Let $(\m,g)$ and $(\m',g')$ be two past- and\linebreak future-distin\-gui\-sh\-ing spacetimes, each supplied with a metric connection to inherit a Riemannian structure, and let $f$ be a causal isomorphism (i.e., keeping the chronological relation as well as its inverse) from $\m$ to $\m'$. Then, $f$ is a smooth conformal diffeomorphism. Furthermore, $f$ preserves timelike curve orientations.}
\end{quotation}
By definition, $f$ is a conformal diffeomorphism preserving orientations if $f^*(g')=e^{2v}\,g$, where $v$ is a continuous function defined on $\m$. If $v=0$, then $f$ is said to be isometric.
\section{A solution to the clock paradox using conformal proper times\label{coeur}} 
\subsection{The geometric settings}
In this preliminary section, we present the general mathematical framework used in the next two sub-sections treating the clock paradox. 
\par
Let $\m$ be a Woodhouse spacetime supplied with a Lorentzian metric $g_0$ (defined from a scalar potential of metric) and a metric connection denoted by $\nabla^0$ providing $\m$ with a Riemannian structure. We assume $g_0$ is at least of class $C^1$ and that it is so defined from clock and message functions at least of class $C^3$ as well as $\m$.
Then, we consider two intersecting particles (or \textit{`trips'}, following the Woodhouse definition) $p$ and $\tp$. We point out again that the non-intersection is only considered in the definition of the $C^2$-congruences, and so it is associated with the choice for the projective structure defined by $\nabla^0$ but not with the whole set of particles or in particular, $p$ and $\tp$. In other words, there are two other $C^2$-congruences, each containing $p$ or $\tp$ and each associated with another possibly non-equivalent projective structure.
We denote by $U_\tp$ a tubular past- and future-reflecting open neighborhood of $\tp$, onto which the message functions $f^\pm_\tp:U_\tp\longrightarrow\tp$ are defined. The neighborhood $U_\tp$ should possibly be reduced to avoid including caustics. 
\par
In addition, $p$ and $\tp$ intersect at two points $o$ and $\iota$ only, the latter being contained in a past- and future-reflecting open sub-neighborhood $V_\tp$ of $U_\tp$. Moreover, we assume that $p$ and $\tp$ are contained in $V_\tp$ between these two points (chronologically). The same notation $f^\pm_\tp$ is used for the restrictions of $f^\pm_\tp$ to $V_\tp$. We set $V\equiv V_p\cap V_\tp\not=\emptyset$, where $V_p$ is analogously defined to $V_\tp$. In order to apply the Malament Theorem, $V$ is assumed to be  connected and without boundaries. Moreover, we also consider $V$ as a conformal manifold, i.e., a manifold being endowed with a class $[g_0]$ of conformally equivalent Lorentzian metrics $g$ of which $g_0$ is a representative and being supplied with its natural causal structure, i.e., we use the natural horismos $\to$. In addition, the class $[\nabla]$ of metric connections $\nabla$ considered throughout, each supplying $V$ with a Riemannian structure compatible with the conformal structure defined by $g$ (or $g_0$), represents those metric connections such that, at least, together $p$ \textit{and} $\tp$ are projective geodesics between $o$ and $\iota$. Then, only on $p$ and $\tp$ are these specific \textit{restricted} Weyl structures defined.
\par\smallskip
If $t'_p:p\longrightarrow\Rset$ and $t_\tp:\tp\longrightarrow\Rset$ are clock functions, respectively, on $p$ and $\tp$, then $p$ can also be parameterized with $t_\tp$ using $f^+_p$. Indeed, we can set $\forall x^+\in p$: $t'_p(x^+)=t_\tp(\tx)$, where $x^+=f^+_p(\tx)$. We choose $t_\tp$ such that 
$t_\tp(o)=0$ and $t_\tp(\iota)=1$.
\par\smallskip
Let $\xi'$ and $\txi$ be the vector fields tangent, respectively, to $p$ and $\tp$ and defined from the parameterizations $t'_p$ and $t_\tp$.
Then, in full generality, we have on $V$: $g_0(\xi',\xi'{)}\equiv e^{2\alpha'}$ and $g_0(\txi,\txi\,)\equiv e^{2\tilde{\alpha}}$, where $\alpha'$ and $\tilde\alpha$ are real differentiable functions defined, respectively, on $W_p\equiv p\cap V$ and $W_\tp\equiv\tp\cap V$ only.
Let $g\equiv e^{-2\chi}g_0$ be a metric field defined on $V$, where $\chi$ is a real differentiable function on $V$ such that $\chi(x)\equiv\alpha'(x)$ if $x\in W_p$ and $\chi(\tx)\equiv\tilde\alpha(\tx)$ if $\tx\in W_\tp$. Thus, the relations $g(\xi',\xi')=1$ and $g(\txi,\txi)=1$ hold, and $g\in[g_0]$.
\par\smallskip
Thenceforth, we can consider the parallel transport from $x\in W_p$ to $\tx^+\equiv f^+_\tp(x)$ along a piecewise $C^3$ null geodesic $L$ (geodesic with respect to both the conformally equivalent metrics $g_0$ and $g$) in $\partial\bar{I}^{+}(x)\cap\partial\bar{I}^{-}(\tx^+)\cap V$. We denote by $k$ the vector field tangent to $L$, and by $\nabla$ a metric connection associated with $g$, as previously indicated.
Then,  we define the vectors $\txi^{+}_\lambda$ ($\lambda\in[0,1]$) parallel transported from $\xi'(x)$ along $L$, such that $\txi^{+}_0=\xi'(x)$ and $\nabla_k\,\txi^{+}_\lambda=0$, where $\nabla_k$ is the covariant derivative in the direction of $k$. The transported vectors are reached using the parallel transport map $\Gamma_\lambda(L)$ which is defined such that:
\begin{equation}
\Gamma_\lambda(L): \eta \in{}T_x\m\longrightarrow\tilde\eta^+_\lambda\in{}T_{\ell(\lambda)}\m
\end{equation}
where $\ell(\lambda)\in L$, $\ell(0)=x$ and $\ell(1)=\tx^+$, and $\ell$ is continuous and increasing for the horismos. On the other hand,
this is a fundamental map accounting for the causal interpretations that each observer makes locally from a signal being propagated along a null geodesic. Obviously, from the vanishing covariant derivative of  $\txi^+_\lambda$  in the direction of $k$, we have along $L$, $g(\txi^+_\lambda,\txi^+_\lambda)=1$.
\par\smallskip
Let $dt'_p$ and $dt_\tp$ be the dual one-forms defined, respectively, on $p$ and $\tp$ such that $dt'_p(\xi')=dt_\tp(\txi)=1$. Then, we define on $T^*\tp$ the  one-form $\tOm$  such that:
\begin{equation}
\tOm\equiv\frac{dt_\tp}{g(\txi^+_1,\txi)}.
\label{tO}
\end{equation}
\begin{figure}[ht]
\begin{center}
\includegraphics[scale=.4]{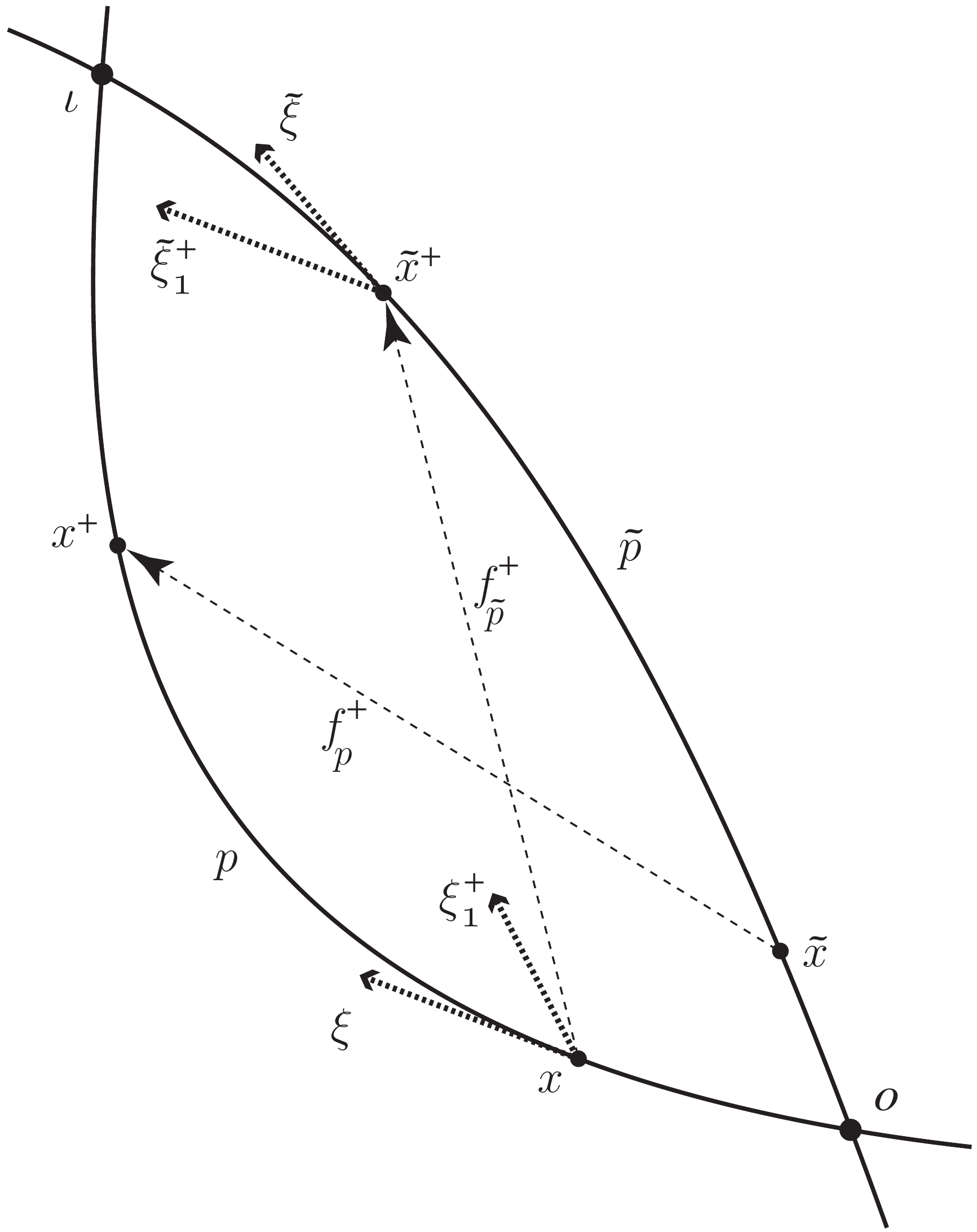}
\par
Figure 1
\end{center}
\end{figure}
We now consider the inverse situation with two other parameterizations for the particles $p$ and $\tp$, namely, $t_p$ and $t'_\tp$ respectively, but using instead the map $f_\tp^+:U_\tp\longrightarrow\tp$ (see Figure~1). In the same way, we set $\forall \tx^+\in\tp$: $t'_\tp(\tx^+)=t_p(x)$ where $\tx^+\equiv f^+_\tp(x)$, and $t_p(o)=0$ and $t_p(\iota)=1$. Thus, one obtains an expression analogous to \eqref{tO} along $\tilde L$, where 
$\xi$ and $\xi^+_1$ are defined in a way similar 
to $\txi$ and $\txi^+_1$:
\begin{equation}
\Omega\equiv\frac{dt_p}{g(\xi^+_1,\xi)}.
\label{O}
\end{equation}
In conclusion, only the parameterizations $t_p$ on $p$ and $t_\tp$ on $\tp$ are relevant, with the others being deduced afterwards from the message functions.
\par\medskip
Lastly, if $V$ is relatively compact, the integrals computed subsequently will always be defined, since the integrands will be Lipschitzian. However, in addition, the Lipschitz constants will no longer depend on $V$, according to the Zeghib Theorem \citeyearpar{zeghib2004}. Indeed, all the maps on $V$ are defined from a foliation of codimension one defined from a given $C^2$-congruence.
\subsection{The non-resolution of the clock paradox induced by simultaneity and proper times}
In the present framework, we shall define a simultaneity map and certain types of proper times from which the first difficulty that arises is the realization of a differential aging formula consistent with the experimental results.
\par
Let the set of pairs $(x,\tx)$ be such that $t_p(x)=t_\tp(\tx)$.
An expression developed for $t_p(x^+)$ could be: $t_p(x^+)\equiv t_p\circ f^+_p(\tx)=t_p\circ f^+_p\circ t_\tp^{-1}\circ t_p\circ f^-_p(\tx^+)$, and thus, $t_p(x^+)\neq t_\tp(\tx^+)$ a priori. Setting this equality is like defining $t_p$ as a function of $t_\tp$, or conversely. Then, we impose the condition $t_p(x^+)=t_\tp(\tx^+)$.
\par
Now, let the map $F:p\longrightarrow\tp$ be such that $F(x)=\tx$ when
$t_p(x)=t_\tp(\tx)$. Thus, the relation $F(x^+)=\tx^+$ necessarily holds, since $t_p(x^+)=t_\tp(\tx^+)$. Moreover, $F$ is bijective and reversible. The map $F$ transforms timelike curves into other timelike curves. However, if $x\ll y$ on $p$, then on $\tp$, it is equivalent to $t_p(x)< t_p(y)$ $\Longleftrightarrow$ $t_\tp(\tx)< t_\tp(\tilde y)$
$\Longleftrightarrow$ $\tx\ll\tilde y$ $\Longleftrightarrow$ $F(x)\ll F(y)$. Thus, the map $F$ is a causal isomorphism from $p$ to $\tp$.
However, from the Malament Theorem, there exists  on $V$ a (non-unique) $C^3$ conformal extension $\wf$ of $F$, which, in addition, preserves the particle orientations while passing from $p$ to $\tp$.
\par\smallskip
Let us note that the map $F$ can never be one of the message functions $f^\pm_\tp$ restricted to $p$, because in this case, the relation $\tx=\tx^+$ with $x\ll x^+$ should hold with $F$ being causal, which is obviously impossible. One calls $F$ the `\textit{simultaneity map}.' It is almost fully geometrically unspecified a priori and is non-unique, depending on the parameterizing clock functions. However, it is well-known that, in contrast, it is causally worked out as being unique, as shown with the Malament Theorem on simultaneity \citep{malam77bis}.
\par\smallskip
From this point on, we shall define a variant of the chronometric hypothesis that is nonetheless quite different. Instead of assuming that the interval $ds$ associated with $g$ is identified as $ds\equiv{c}\,d\tau$, where $\tau$ is the proper time of a clock, we suppose that $ds\equiv{c_p}\,dt_p$ on $p$, and $ds\equiv c_\tp\,dt_\tp$ on $\tp$. That is to say, the interval $ds$ is identified as the sole ``physically" measurable quantity, which represents the time intervals $dt_p$ \textit{displayed} by the clocks.
Therefore, these are as varying as the clocks used, contrary to the proper times and depending on each particle but not on their parameterizations. Moreover, the constants $c_p$ and $c_\tp$ are the numerical values for the speed of light when making use of the corresponding parameterizations $t_p$ and $t_\tp$. Then, we set:
\begin{equation}
\tOm\equiv\delta\tau_\tp,
\qquad
\Omega\equiv\delta\ttau_p,
\end{equation}
where $\tau_\tp$ and $\tilde\tau_p$ are the ``usual" proper times associated with (not defined on) the particles $p$ and $\tp$, respectively.
The one-form $\tOm$ is ascribed to the infinitesimal proper time $\delta\tau_\tp$ of $p$, because, first, $\txi^+_1$ depends on two points, one on $p$ and the other on $\tp$, and secondly, the term $g(\txi^+_1,\txi)$ is the well-known $\gamma$ factor of special relativity between one observer on $p$ and another on $\tp$. Hence, the proper time of an object on $p$ is evaluated on $\tp$, onto which $\tOm$ is defined, hence the notation. Implicitly, this shows the ``relative" character of the concept of proper time, which will also be truly justified further by using certain other arguments presented below.
Then, these two one-forms are defined strictly on the product $p\underset{F}{\times}\tp\,$ ``fibered" by a simultaneity map $F$.
\par
In addition, since $\wf$ puts into correspondence the (co)tangent vector spaces on $p$ and $\tp$, the relations 
$\wf^*(dt_\tp)=dt_p$ and $\wf_*(\xi)=\txi\,$ hold  while preserving the particle orientations, since $t_p(x)=t_\tp\circ F(x)$.
Moreover, at least between the points $o$ and $\iota$, we have the important property $\wf(p)=\tp$, but naturally $\wf(\tp)\neq p$ a priori.
We suppose this last equality is satisfied when choosing, among all of the maps $\wf$, those that satisfy the condition $C_1$ below.
\begin{definition} Let $\wf$ be a conformal diffeomorphism, such that $\wf(p)=\tp$.
We denote by $C_1$ the condition $\wf^2=\mathbb{I}d$ when $\wf$ is restricted to $p$ or $\tp$ only.
\end{definition}
Therefore, $\wf$ carries out the exchange of the two particles.
From all of these relations, we have $\wf^*(\tOm)=\Omega$ on $p$ and $\tp$, only if $\wf^*(g)(\wf_*^{-1}(\txi^+_1),\xi)=g(\xi^+_1,\xi)$.
This relation would be satisfied if, in particular and due to the Malament Theorem, the following equalities
$\wf_*^{-1}(\txi^+_1)=e^r\,\xi^+_1$ and $\wf^*(g)=e^{-r}\,g$
held on $p$.
In fact, we would obtain the relation $\wf^*(g)=g$ on $p$, because $\txi^+_1$ and $\xi^+_1$, as well as $\xi$ and $\txi$, have the same norms with respect to $g$. 
In order to satisfy these equalities, we set below the defining constraint or hypothesis on $\wf$, namely, the constraint $C_2$:
\begin{definition} $\wf$ satisfies the condition $C_2$ if
$\wf$ preserves the Riemannian structure on $(V,g)$.
\end{definition}
Then, as we will show in the proof below, the relation $\wf_*^{-1}(\txi^+_1)=\xi^+_1$ holds if $C_1$ and $C_2$ are assumed. 
Then, we have:
\par\medskip\noindent
\textbf{Lemma}
\textit{-- If $C_1$ and $C_2$ are satisfied, then 
$\wf^*(\tOm)=\Omega$ on $p$ and $\tp$.}
\proof
Indeed, firstly, if $\wf$ preserves the Riemannian structure, then for all continuous vector fields $\mu$ and $\zeta$ on $V$, there exists a continuous function $w$ on $V$ such that $\wf_*(\nabla_\mu\zeta)=e^{2\,w}\,\nabla_{\mu'}\zeta'$,
where $\mu'=\wf_*(\mu)$ and $\zeta'=\wf_*(\zeta)$. Then, considering $k$, a null vector tangent to the piecewise $C^3$ null geodesic $L$ from $x$ to $\tx^+$, and $\txi^+_\lambda$ ($\lambda\in[0,1]$), the parallel transported vectors along $L$ of $\xi$ at $x$, we find $\nabla_{k'}\xi'^+_\lambda=0$, where $k'=\wf_*(k)$ and $\xi'^+_\lambda=\wf_*(\txi^+_\lambda)$.
Secondly, since $x\to\tx^+$ then $F(x)=\tx\to F(\tx^+)=x^+$. Therefore, from Lemma 4.2 of Woodhouse, the  geodesic $L$ from $x$ to $\tx^+$ is unique as well as a piecewise $C^3$ null geodesic from $\tx$ to $x^+$ denoted by $\widetilde{L}$. Moreover,  the latter is such that $\wf(L)=\widetilde{L}$ on $V$. Moreover, since $\xi'^+_0=\xi^+_0=\txi$ at $\tx$ and $\widetilde{L}$ is unique, then for all $\lambda\in[0,1]$, we find  $\xi'^+_\lambda=\xi^+_\lambda$ on $\widetilde{L}$. Thus, $\wf_*(\txi^+_1)=\xi^+_1$ and $\wf^*(g)=g$. However, from $C_1$, we also have $\wf_*^{-1}=\wf_*$ when $\wf$ is restricted on $\tp$ and so the expected result. Then, $\Omega$ is the pull-back of $\tOm$ by $\wf$.
\hfill$\Box$
\par\smallskip
Hence, on the basis of a physical interpretation of the signals received by each particle, i.e., considering again that the proper time of each particle is actually assessed as such only by another particle running in the ``time" given by its own clock (in special relativity, the proper time for a physical frame is always compared, using a factor $\gamma$, to another physical frame), then the differential aging $\triangle^0_{(\tp,p)}(\tp,p)$ between the clocks can be computed \textit{a priori} with the formula:
\begin{definition} We set $\triangle^0_{(\tp,p)}(\tp,p)$, the differential aging, with the ``usual" proper times:
\begin{equation}
\triangle^0_{(\tp,p)}(\tp,p)\equiv\int_o^\iota \delta\ttau_p\,\,-\int_o^\iota \delta\tau_\tp=\int_o^\iota \frac{dt_p}{g(\xi^+_1,\xi)}-\int_o^\iota \frac{dt_\tp}{g(\txi^+_1,\txi)}\,.
\label{fdiff}
\end{equation}
\end{definition}
Then, in making use of $\wf$ and the constraints $C_1$ and $C_2$, we find:
\begin{equation}
\triangle^0_{(\tp,p)}(\tp,p)=\int_o^\iota (1-1)\,\frac{dt_p}{g(\xi^+_1,\xi)}=0\,.
\label{exptau}
\end{equation}
Thus, we must clarify the links between the constraints on $\wf$ and some physical properties. Nevertheless, we cannot make allowance for concepts inherited from special relativity to define a differential aging formula as above.
\par\smallskip
Now, in order to seek and to justify a correct formula for the differential aging, we will also consider a situation with a third particle $q$, which is more related to physical experiments that have actually been performed, such as the Hafele and Keating experiment~\citeyearpar{hafelekeat172,hafelekeat272}.
\subsection{Conformal proper times and differential aging deduced from a third particle independence}
Let $q$ be a third particle not necessarily passing through the points $o$ or $\iota$, with its two message functions~$f^\pm_q$. We assume that $f^+_q(o)$ and $f^+_q(\iota)$ as well as every point $x$ of $q$ such that $f^+_q(o)\ll x\ll f^+_q(\iota)$ are in $V$.
According to $q$ and the signals it receives coming from $p$ and $\tp$, i.e. via the parallel transports along light paths, a difference of proper times
$\triangle^0_{(q,q)}(\tp,p)$ between the two particles $p$ and $\tp$ could be written as follows:
\begin{definition} We define
\begin{equation}
\triangle^0_{(q,q)}(\tp,p)\equiv\int_{f^+_q(o)}^{f^+_q(\iota)}
\left\{
\frac{1}{g(\eta,\tzeta^+_1)}-\frac{1}{g(\eta,\zeta^+_1)}
\right\}
dt_q,
\label{diffq}
\end{equation}
where $\zeta^+_1$ and $\tzeta^+_1$ are the parallel transported vectors at $y=f^+_q(x')=f^+_q(\tx)\in q$ from, respectively, the vectors $\xi'\in{}T_{x'}p$, $\txi\in{}T_{\tx}\tp$, and $\eta$ is a vector field tangent to $q$ provided with the parameterization $t_q$  (see Figure 2).
\end{definition}
\begin{figure}[ht]
\begin{center}
\includegraphics[scale=.4]{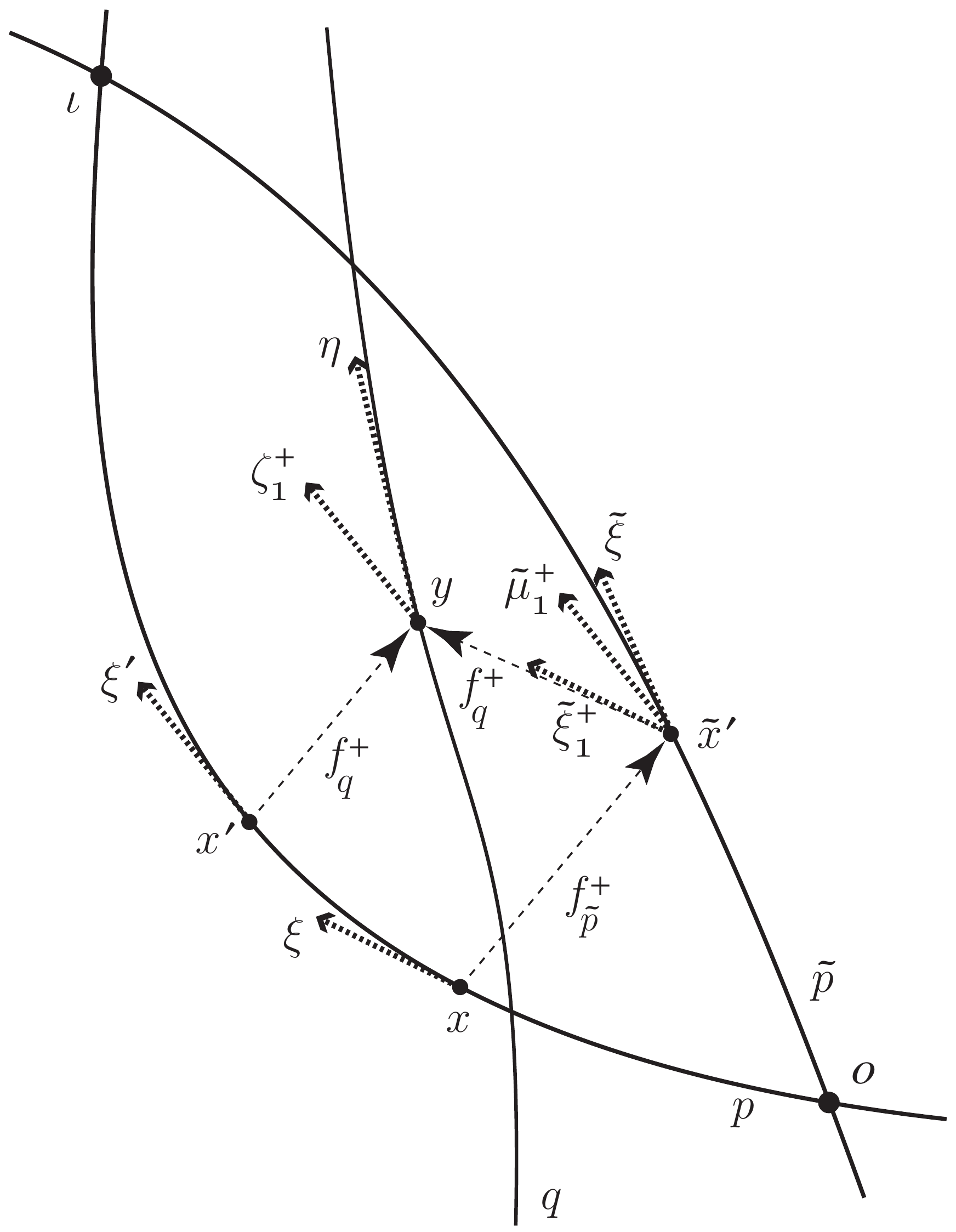}
\par\bigskip
Figure 2
\end{center}
\end{figure}
In addition, we assume that $g=e^{-2\,\chi}\,g_0$ with $\chi=\nu$ on $q$, where $g_0(\eta,\eta)=e^{2\,\nu}$. Moreover, we set $r^+_q(o)=0$ and $r^+_q(\iota)=1$. The expression \eqref{diffq} must be absolutely independent on $q$, and we shall show that a particular condition must be satisfied to obtain such independence.
\par \smallskip
We denote by $L'$ the null geodesic from $x'\in p$ to $y\in q$ (see Figure 2), $\widetilde{L}$ from $\tx\in\tp$ to $y$, and $L$ from $x\in p$ to $\tx\in\tp$. In addition, we denote again by $\eta$ the vector field tangent to $q$ at $y=f^+_q(x')=f^+_q(\tx)$ and $\zeta^+_1$ and $\txi^+_1$, the parallel transported vectors, such that:
\begin{equation}
\zeta^+_1=\Gamma_1(L')(\xi'),
\qquad
\txi^+_1=\Gamma_1(L)(\xi).
\end{equation}
Then, we have:
\par\medskip
\noindent\textbf{Proposition.} 
\textit{There exists a unique differentiable function $\beta$ on $p$, independent on the representative metric connection $\nabla$ and $q$, such that $\beta(o)=0$ and
\begin{equation}
\frac{e^{-2{\beta}(x')}}{g(\zeta^+_1,\eta)\circ{}f^+_q}\equiv\frac{e^{-2{\beta}(x)}}{g(\txi^+_1,\txi)},
\label{propint}
\end{equation}
at $\tx\in\tp$ with $x=f^-_p(\tx)$ and $x'=f^-_p\circ f^+_q(\tx)$.
We obtain an analogous relation for $\tilde\beta$ on~$\tp$.}
\par
\proof Let $f^-_{\tp/q}$ be the message function $f^-_{\tp}$ restricted to $q$.
As for $F$, we extend the latter map to a $C^3$ causal conformal map defined on $V$ and denoted by $\hf^-$.
The relation $\hf^-_*(\eta)=\txi$ holds, and since $\eta$ and $\txi$ have the same norms with respect to $g$, $(\hf^-)^*(g)=g$. Additionally, we define $\tmu^+_1$ such that \(\hf^-_*(\zeta^+_1)\equiv\tmu^+_1\).
Then, we have the relations:
\begin{equation}
g(\zeta^+_1,\eta)\circ f^+_q(\tx)=g(\tmu^+_1,\txi)\neq g(\txi^+_1,\txi).
\label{pass}
\end{equation}
However, $\hf^-_*$ acts as an isometry with respect to $g$ on the vectors
$\eta$ and $\zeta^+_1$, as well as $\Gamma_1(L')$ on $\xi'$. Thus, we can set the relations:
 \begin{equation}
\TL'_g(\xi')\equiv\hf^-_*\circ\Gamma_1(L')(\xi')=\tmu^+_1,
\qquad
\TL_g(\xi)\equiv\Gamma_1(L)(\xi)=\txi^+_{1},
\end{equation}
where $\TL'_g$	and $\TL_g$ are isometries with respect to $g$.
On the particle $p$, the carriage of $\xi$ from $x$ to $x'$ is not a parallel transport with respect to $g_0$, because $p$ is not a priori geodesic under $g_0$. Accordingly, we can set
\(\nabla^0_\xi\xi\equiv f_0\),
although, $\nabla_\xi\xi=0$. Setting $\mu=e^\rho\xi$, if 
$\nabla^0_\xi(e^\rho\xi)=e^\rho(d\rho(\xi)\,\xi+f_0)=0$, then we necessarily  find: $d\rho(\xi)+g(f_0,\xi)=0\Longleftrightarrow d\rho+g(f_0,\xi)\,dt_p=0$ on $p$. Then, integrating from $x$ to $x'$, we deduce the integral definition of $\rho$:
\begin{equation}
\rho(x')=-\int_x^{x'}g(f_0,\xi)\,dt_p\,.
\end{equation}
From this definition, we set $\rho(x')\equiv\beta(x')-\beta(x)$, where
\begin{equation}
\beta(x)=-\int_o^{x}g(f_0,\xi)\,dt_p\,.
\end{equation}
Additionally, $g_0(\mu,\mu)=cst$, since $\nabla^0_\xi\mu=0$, but
$\mu=\xi$ at $x$ and thus, $g_0(\mu,\mu)=1$. In conclusion, there exists an isometric transformation $\TL_{g_0}$ with respect to $g_0$, such that $\mu=\TL_{g_0}(\xi)$, and consequently:
\begin{equation}
\xi'=e^{\beta(x)-\beta(x')}\,\TL_{g_0}(\xi).
\label{tlg0}
\end{equation}
Thus, we deduce with $\Lambda\equiv\TL'_g\circ\TL_{g_0}\circ\TL_g^{-1}$ that:
\begin{equation}
\Lambda(\txi^+_{1})\equiv
e^{\beta(x')-\beta(x)}\,\tmu^+_{1}\,.
\label{monodr}
\end{equation}
However, $\txi^+_1$ and $\tmu^+_1$ are normalized with respect to the metric $g$ at $\tx$, namely, the metric we denote hereafter by $h$. 
Thus, necessarily, the holonomy map along the loop $\ell\equiv(\tx x \,x' y\, \tx)$, i.e., $\Lambda$, is a dilatation composed with a Lorentz transformation $\TL_h$ of $h$: 
\begin{equation}
\Lambda\equiv e^{\beta(x')-\beta(x)}\,\TL_h,
\end{equation}
such that $\TL_h(\txi^+_1)=\tmu^+_1$.
The map $\Lambda$ can always be composed on the left with another Lorentz transformation $\TL'_h$ preserving $\tmu^+_1$, but such that $\TL'_h\circ\TL_h(\txi)=\txi$. Therefore, in full generality, we can always choose a map $\Lambda$ such that the associated Lorentz transformation $\TL_h$ remains $\txi$ invariant. Thus, we deduce that there exists a map $\Lambda$ and, moreover, a function $u$ on $\tp$ such that: 
\begin{equation}
h(\Lambda(\txi^+_1),\Lambda(\txi))=e^{2(\beta(x')-\beta(x))}\,h(\tmu^+_1,\txi)
=e^{2(\beta(x')-\beta(x)+u)}\,h(\txi^+_1,\txi).
\label{holon}
\end{equation}
Therefore, in relation \eqref{pass}, the only way to pass from $\txi^+_1$ to $\tmu^+_1$ using signaling, that is to say, passing from 
$p$ to $\tp$ using null geodesics, is to apply $\Lambda$.  However, on the loop $\ell$, equivalently, the signals at $y$ coming 
from $x$ sometimes via $x'$ or sometimes via $\tx$ must be the same. This signal identification, regardless of the light paths followed, means that $\Lambda$ leaves $h$ invariant. Hence, the relation $u=\beta(x)-\beta(x')$ must hold. Additionally, this means that the holonomy group of $\nabla$ at each point in $V$ is the Lorentz group of $g$ at this point. This is the present case, since, from condition $C_2$, the metric connection $\nabla$ provides $V$ with a Riemannian structure \citep{ehl72}. Thus, there always exists a map $\Lambda$ leaving $h$ invariant. Then, at the point $\tx$ and with 
$f^+_q$ restricted to $\tp$, we find:
\begin{equation}
\frac{1}{g(\zeta^+_1,\eta)\circ f^+_q}=e^{2(\beta(x')-\beta(x))}\,\frac{1}{g(\txi^+_1,\txi)},
\end{equation}
with $x=f^-_p(\tx)$ et $x'=f^-_p\circ f^+_q(\tx)$. The relation \eqref{propint} holds. In addition, we can deduce at $x\in p$, using an analogous transformation $\widetilde\Lambda$ and with $f^+_q$ restricted to $p$:
\begin{equation}
\frac{1}{g(\tzeta^+_1,\eta)\circ f^+_q}=e^{2(\tilde{\beta}(\tx')-\tilde{\beta}(\tx))}\,\frac{1}{g(\xi^+_1,\xi)}\,,
\end{equation}
with $\tx=f^-_\tp(x)$ and $\tx'=f^-_\tp\circ f^+_q(x)$.\hfill$\Box$
\par\medskip
Then, we obtain in particular:
\begin{equation}
\int_{f^+_q(o)}^{f^+_q(\iota)}e^{-2{\beta}(x')}\,\frac{dt_q}{g(\zeta^+_1,\eta)}\equiv\int_o^\iota e^{-2{\beta}(x)}\,\frac{dt_\tp}{g(\txi^+_1,\txi)},
\end{equation}
with $x=f^-_p(\tx)$ and $x'=f^-_p(y)$. 
However, $q$ may vary and be such that $q=p$. The relation 
\begin{equation}
\int_{o}^{\iota}e^{-2{\beta}}\,dt_p\equiv\int_o^\iota e^{-2{\beta}\circ f^-_p}\,\frac{dt_\tp}{g(\txi^+_1,\txi)}
\label{expbet}
\end{equation}
holds as well. Performing a change of variables using the map $\wf$, we find:
\begin{equation}
\int_{o}^{\iota}e^{-2{\beta}}\,dt_p\equiv\int_o^\iota e^{-2{\beta}\circ f^-_p\circ F}\,\frac{dt_p}{g(\xi^+_1,\xi)}.
\end{equation}
However, the two integrands are positive, so we can set new supplementary defining constraints $C_3$ and $C_4$ on $\wf$, which is not fully defined since on $p$, if $x\ll x''$, then we find only: $\forall x'\in p$ such that $x\ll x'\ll x''$, then $F(x')\in\,\,]\,f^-_\tp(x),f^+_\tp(x'')\,[\,\,\subset\tp$. 
By definition, the constraints $C_3$ and $C_4$ are:
\begin{definition} $F$ satisfies the condition $C_3$ if 
\begin{equation}
e^{-2\beta}=\frac{e^{-2\beta^-_p\circ F}}{g(\xi^+_1,\xi)}\,,\qquad\beta^-_p\equiv\beta\circ f^-_p,
\end{equation}
and the condition $C_4$ if
\begin{equation}
e^{-2\tilde\beta}=\frac{e^{-2\tilde\beta^-_\tp\circ F}}{g(\txi^+_1,\txi)}\,,\qquad\tilde\beta^-_\tp\equiv\tilde\beta\circ f^-_\tp.
\end{equation}
\end{definition}
Nevertheless, we must examine whether any of these constraints are redundant.
However, setting $x_-=f^-_p\circ F(x)$ and $\tx_-=f^-_\tp\circ F(\tx)$, then necessarily, $F(x)=\tx$ and $F(x_-)=\tx_-$\,. Hence, a constraint is equivalent to the other if and only if $\tilde\beta\circ F=\beta$, a condition that is not always satisfied, and therefore, these constraints are not equivalent. In addition, we point out that these two constraints could be ``physical" constraints rather than ``geometrical" constraints if the functions $\beta$ are ascribed to physical quantities, and that is indeed the case, as we shall see. Then, the simultaneity map $F$ is related to the physics instead of the causality only.
\par\smallskip
Thus, we can conclude that $\triangle^0_{(q,q)}(\tp,p)\neq\triangle^0_{(\tp,p)}(\tp,p)\equiv0$. Then, although the exponential factors in the integrands may have common constant factors, we set a priori for all $q$ the definitions below for the differential aging between $p$ and $\tp$:
\begin{definition} We set the differentials $\triangle_{(\tp,p)}(\tp,p)$ and $\triangle_{(q,q)}(\tp,p)$  such that:
\begin{equation}
\triangle_{(\tp,p)}(\tp,p)\equiv
\int_o^\iota e^{-2\tilde\beta^-_\tp}\,\frac{dt_p}{g(\xi^+_1,\xi)}
-\int_o^\iota e^{-2\beta^-_p}\,\frac{dt_\tp}{g(\txi^+_1,\txi)},
\label{twin}
\end{equation}
and
\begin{equation}
\triangle_{(q,q)}(\tp,p)\equiv
\int_{f^+_q(o)}^{f^+_q(\iota)}
\left\{
\frac{e^{-2\tilde\beta^-_\tp}}{g(\eta,\tzeta^+_1)}-\frac{e^{-2\beta^-_p}}{g(\eta,\zeta^+_1)}
\right\}
\,dt_q.
\label{diffet}
\end{equation}
\end{definition}
From these new definitions and the previous developments, we have:
\par\bigskip\noindent
\textbf{Theorem.}
\textit{If the conditions $C_1$ to $C_4$ are satisfied on $V$, then
\begin{equation}
\triangle_{(q,q)}(\tp,p)=\triangle_{(\tp,p)}(\tp,p),
\end{equation}
regardless of the particle $q$.}
\proof Obvious from the Proposition and the previous definitions.\hfill$\Box$
\par\medskip
Then, $\triangle_{(q,q)}(\tp,p)$ is invariant with respect to $q$.
Nevertheless, the proper times must be redefined. 
\begin{definition}
The conformal (and relative) proper times are defined by the relations:
\begin{equation}
\delta\ttau_p\equiv e^{-2\tilde\beta^-_\tp}\,\frac{dt_p}{g(\xi^+_1,\xi)}\,,\qquad \delta\tau_\tp\equiv e^{-2\beta^-_p}\,\frac{dt_\tp}{g(\txi^+_1,\txi)},
\end{equation}
and
\begin{equation}
\delta\tau_q\equiv e^{-2\beta^-_p}\,\frac{dt_q}{g(\eta,\zeta^+_1)}\,,\qquad \delta\ttau_q\equiv e^{-2\tilde\beta^-_\tp}\,\frac{dt_q}{g(\eta,\tzeta^+_1)}.
\end{equation}
\end{definition}
\par\smallskip
Two other formulas can be given for the expression \eqref{twin} of the difference of proper times. The first is obtained with $C_3$ or \eqref{expbet}:
\begin{equation}
\triangle_{(p,p)}(\tp,p)=\int_o^\iota \left\{
\frac{e^{-2\tilde\beta^-_\tp}}{g(\xi^+_1,\xi)}
-e^{-2\beta}\right\}\,dt_p,
\label{twinp}
\end{equation}
and the second on $\tp$, analogous to the first, with the condition $C_4$. We note that they are close to \eqref{diffet} and they indeed reduce to \eqref{diffet} if, for example, $p=q$ in \eqref{twinp}.
\par\smallskip
The relation \eqref{diffet} allows us to solve the clock paradox, since the unspecified particle $q$ can be taken as equal to $p$ or $\tp$.
However, this amounts to modifying the computation viewpoints for the integral of the differential aging formula. By ``viewpoint", we mean that the integrating variable specifies it, i.e., the pairs $(q,q)$ or $(\tp,p)$ in the subindices in the differences $\triangle(\tp,p)$. 
As a result, we see that the computation of the difference of times between the clocks is the same as that evaluated by the two clocks:
\begin{equation}
\triangle_{(q,q)}(\tp,p)=\triangle_{(\tp,p)}(\tp,p)=\triangle_{(\tp,\tp)}(\tp,p)=\triangle_{(p,p)}(\tp,p).
\end{equation}
Finally, the chronometric hypothesis must be amended by the conformal factors and the ``gamma" factors.
For this, we consider the particular viewpoint in the Minkowski spacetime with $p=q$. It follows that $\eta=\zeta^+_1=\xi$ and (considering that the functions $\beta$ vanish for simplicity):
\begin{equation}
\triangle_{(q,q)}(\tp,q)=\int_{o}^{\iota}
\left\{
\frac{1}{g(\xi,\tzeta^+_1)}-1
\right\}
\,dt_q,
\end{equation}
and thus finally, if $g$ is close to $g_0$, i.e., $t_p$ is close to the proper time of the particle $p$, the usual factors $\gamma$ obtained from the Minkowski metric can be used. Therefore, we find:
\begin{equation}
\triangle_{(q,q)}(\tp,q)\equiv 
\int_{o}^{\iota}\tfrac{dt_q}{\gamma}-
\int_{o}^{\iota}dt_q\,<0,
\end{equation}
which is clearly the same formula that is usually used in the Minkowski spacetime.
However, we also identify the origin of the CH, by taking for a proper time a very particular clock function $t_q$, which is substituted, due to the CH, by a known gravitational time. However, the question then follows: if a clock stops beating its unit of time and remains in a physical ``steady state", does its proper time change? Meaning, is its proper time what the clock displays? The reply might be yes (!), depending on which other ``times" the clock sequentially displays as its ``own" displayed time $t_q$.
The conclusion is essential in itself: the functions $\beta$ make it possible to connect the data of a temporal counter, like a clock with $t_q$, to a physical duration. Thus, necessarily, the previous functions are physical and associated mainly with a relation between a physical state of an object with its information content and other objects with their own physical states and information content.
\subsection{The physical interpretations of the conformal factors}
The physical/geometrical interpretations of the functions $\beta$, for instance, cannot really be performed without the use of some results given by typical physical experiences in general relativity. Nevertheless, they can be related to more general physical aspects. We propose this as a first step.
\par\smallskip
First of all, the ``natural" hypothesis would be that these functions come out of metrics $g$ with the general form $g\equiv e^{-2w}\,g_0$ defined only on $p$ (or $\tp$). Then, the functions $w$ cannot be scalar fields on $V$ and a fortiori on $\m$. Nevertheless, at a particular $z$ on $p$, the value $w(z)$ determines a set of continuous scalar fields $v$ defined on any tubular neighborhood of $p$ contained in $V$ such that $w(z)=v(z)$. Hence, $w(z)$ defines what is called a germ at $z$, i.e., a class of functions $v$ satisfying the previous equality. We denote by $[w]_z$ this class of functions $v$ and the full set of such classes on $p$ defines a `sheaf' of rings of continuous functions onto $p$, denoted by $S_p(w)\equiv\{[w]_z\,/\,z\in\,p\}$.
\par
Obviously, if $v\in[w]_z$ then we might have $v\not\in[w]_{z'}$ if $z\neq z'$, i.e., $w(z)=v(z)$ and $w(z')\neq v(z')$. Nevertheless, if 
we wish to obtain local results at $z$, we can take any continuous function $v\in[w]_z$ to do the computations. Thus, we can set $g_0\equiv e^{2v}\,g$ at $z$, and use this expression to compute the metric connection $\nabla$ at $z$ associated to $g$ from the metric connection $\nabla^0$ associated with $g_0$. As a result, at $z$ and if $v$ is assumed to be differentiable, the relation:
\begin{equation}
{\nabla}^0_\xi\zeta=\nabla_\xi\zeta+dv(\xi)\,\zeta+dv(\zeta)\,\xi-g(\xi,\zeta)\,\mbox{grad}(v)
\end{equation}
holds, where $\xi$ and $\eta$ are any vector fields defined in an open neighborhood of $z$.
We note that if $v$ and $v'$ both belong to $[w]_z$, then, in general, 
$\mbox{grad}(v)\neq\mbox{grad}(v')$ at $z$. Thus, a priori, there are as many metric connections ${\nabla}$ at $z$ as there are functions in $[w]_z$ whose gradients differ at $z$. Hence, we must define a sub-sheaf of $S_p(w)$ denoted by $S_p(w,f)$ such that $S_p(w,f)=\{[w,f]_z\,/\,z\in\,p\}\subset S_p(w)$, where $f$ is a continuous vector field defined on $p$ only, and the germ $[w,f]_z$ denotes the class of functions $v$ such that $v(z)=w(z)$ and $\mbox{grad}(v)(z)=f(z)$. Any function $f$  only defined on $p$ is suitable (although continuous on $p$), but between $w$ and $f$, we impose the relation at $z\in p$:
\begin{equation}
{\nabla}^0_\xi\zeta=\nabla_\xi\zeta+f^*(\xi)\,\zeta+f^*(\zeta)\,\xi-g(\xi,\zeta)\,f,
\end{equation}
where $f^*$ is the dual one-form of $f$ such that $f^*(\zeta)\equiv g(f,\zeta)$ on $p$. This can be presented with a more meaningful formula.
First, let $\eta$ be any vector field on $p$, with its corresponding space vector $\vec\eta$ such that $\eta\equiv\vec\eta+g(\eta,\xi)\,\xi$, where $\xi$ is assumed to be the timelike vector field tangent to $p$ and normalized with respect to $g$. Of course, $\vec\eta$ satisfies the relations $g(\vec\eta,\xi)=0$ and $g(\vec\eta,\vec\eta)\leq0$. Secondly,  with $\xi$, we have the relation:
\begin{equation}
{\nabla}^0_\xi\xi=\nabla_\xi\xi+2f^*(\xi)\,\xi-f\equiv f_0.
\label{confvec}
\end{equation}
However, since $g(\xi,\xi)=1$, then $g(\xi,\nabla_\xi\xi)=0$, and thus, we
deduce the relations:
\(
\lvec{\nabla_\xi\xi}=\nabla_\xi\xi
\),
and
\begin{equation}
g(\xi,\nabla^0_\xi\xi)=f^*(\xi)\equiv g(\xi,f)=g(\xi,f_0).
\label{force}
\end{equation}
Therefore, we obtain $\nabla^0_\xi\xi=\lvec{\nabla^0_\xi\xi}+f^*(\xi)\,\xi$.
The relation \eqref{confvec} is thus written in the simplified form:
$\lvec{\nabla^0_\xi\xi}=\lvec{\nabla_\xi\xi}-\vec{f}\equiv\vec{f}_0$.
One can rewrite the preceding formula in another form that is more amenable to easy interpretation:
\begin{equation}
\lvec{\nabla^0_\xi\xi}\equiv\vec{f}_0\,,\qquad
\lvec{\nabla_\xi\xi}\equiv\vec{f}_0+\vec{f}.
\end{equation}
These two relations depend on the clock chosen, i.e., the clock function $t_p$ and, thus, the functions $v$. Then, if  there are no applied forces\footnote{We always imply per unit of mass. Thus, it is an applied acceleration rather than an applied force; however, the concept of applied force is more natural than the former, explaining why we now use this abuse of denomination.}  $\vec{f}_0$ when considering the metric $g_0$, i.e., $\vec{f}_0=\vec{0}$, then naturally it is no longer the case considering instead the metric $g$ if $\vec f\neq0$ and, reciprocally, if $\vec f+\vec{f}_0=0$ and $\vec{f}_0\neq0$.
\par
Thus, this exactly represents the `\textit{pseudo-fields}' of the gravity viewpoint, whose explicit terminology can be found in the Einstein paper on the twin paradox \citep{Einstein18,unnik2005} as well as in the lift metaphor of the Einstein equivalence principle \citep{ghinsbuden2001}.
Hence, their origins can be ascribed to the choice of a clock \textit{displaying} a time differing from the proper time. Then, to the lift metaphor, we could append a sort of \textit{unregulated clock metaphor} to special relativity: a massive object is in inertial motion along a straight line at a constant relative speed when compared to a certain clock, and the latter suddenly starts to have a delay that cannot be observed in the absence of an absolute time of reference. As a result, one would observe a sudden increase in the relative speed and, thus, a non-vanishing relative acceleration. 
\par
However, we must mention that the accelerations given by accelerometers remain null, so that the motion remains inertial, even though the relative accelerations are modified. We can ascribe this variation to a geodesic or curvature variation to become non-zero, as well as to a pseudo-field of gravity occurrence. Thus, these fields are ``pseudo" only relative to the choice of the conformal metric carried out a priori during the variation and, thus, of a clock.
\par
If the variations are not considered as being due to changes in temporal coordinates or clocks, then, quite naturally, two metrics $g$ and $g_0$ can be simultaneously associated with each particle $p$. In the geometry defined by $g$, there are no forces applied on $p$, 
i.e., $\vec{f}_0+\vec {f}=0$, and in the geometry defined by $g_0$, a force $\vec{f}_0$ is applied on $p$ and is interpreted as fictive on the metric $g$, i.e., ascribed to a pseudo-field of gravity. On $p$, one would then have, for example, the forces of gravity or centrifugal forces, which would be fictive forces on $g$ but ``true" forces on $g_0$.
\par
Nevertheless, why would that not remain valid if the variations are considered as changes in clocks or clock functions? Indeed, how do we note such a change?
We do not have a local absolute clock against which we can calibrate the others. This is well highlighted in the \textit{Allan deviation}  \citeyearpar{allan66} of a pool of identical clocks. Then, deduced from the latter, the precision of an atomic clock is no longer defined from an absolute, standard clock linked to a given temporal orientation, but from a statistical relation between clocks, and moreover, time drift bias cannot be defined. Thus in practice, an unobservable change in clock function is also equivalent to the sole observed change in geodesics (related to curvature), and then, the lift metaphor can no longer be distinguished from that of an unregulated clock. Therefore, 
this at once poses the question of the relation between clock functions and accelerations, forces or geodesics.
\par\smallskip
Thus, if the relation ${\nabla_\xi\xi}=0$ holds, we should interpret
$g(f_0,\xi)=g(f,\xi)$ as a quantity associated, in some ways, with the mechanical work of a force $\vec{f}_0$, fictive on $g$, or with an energy potential. From the previous equality, one thus notes that the fact that this scalar product does not change when passing from $f_0$ to $f$ means that the work would not be, in any case, fictive. Additionally, we can notice elsewhere that a non-fictive work is at the basis of the Schild argument for deducing the gravitational frequency shifts in light \cite[see pp. 187--189]{misnerwheel73}. Now, starting from the relation 
\eqref{force}, the function $\beta$ should be a priori the mechanical work of force $f$, along $p$ and dependent on the path. As we shall see the function $\beta$ could be the variation in the total energy of the particle minus the kinetic energy.
\par\smallskip
Lastly, it is significant to note that this interpretation for $\beta$ differs for instance from that given by Wheeler 
\citeyearpar{wheeler90} for the Weyl potentials. Wheeler identified $\beta$ with the classical action; it would be very appealing indeed to connect to the physics. However, the examples presented below prevent us, disappointingly, from making such an identification and lead us to dismiss such a proposal in the present situation.
\section{Examples of applications\label{expert}}
One considers in this section only verifications up to the order
$c^2$, corresponding to a first approximation of the definitions of the functions $\beta$, whose refinements in their complete physical interpretations, under study, remain to be made if one wants to move to results up to the order of at least $c^4$.
\subsection{One clock}
This is the simplest situation, occurring when $p=\tp$, but with at least two distinct clocks on the same worldline of a freely falling particle, with one of them not running. Thus, in this case, 
$t_\tp(\tx)=0$ if $\tx$ is on $p$ between $o$ and $\iota$, and it starts to run only while arriving at $\iota$, so that $t_\tp(\iota)=1$.
Then, the relation \eqref{twinp} is written in the form:
\begin{equation}
\triangle_{(p,p)}(\tp,p)=\int_o^\iota(1-e^{-2\beta})\,dt_p=1-\int_o^\iota e^{-2\beta}\,dt_p,
\end{equation}
where $\tilde\beta=0$ because the clock on $\tp$ is not running.
Moreover, $g(f,\xi)$ is then linked only to the internal energy of the running clock, since it is in free fall.
\par\smallskip
Let us consider the term of the entropy $TdS$ in classical thermodynamics, such that $TdS+\delta W\equiv-\,2\,g(f,\xi)\,dt_p$, where $\delta W$ is the infinitesimal ``internal" work (work not related to the kinetic energy). The terms  $S$ and $W$ are dimensionless quantities and represent energies per unit of mass or per unit of thermodynamic energy, such as $kT$ for instance ($k$ being Boltzmann's constant). Setting $S\equiv\mathcal{S}/(k\,T_0)$, where $\mathcal{S}$ is the entropy, and $W\equiv\mathcal{W}/(k\,T_0)$ where $\mathcal{W}$ is the internal work, and assuming $T_0\,dS+\delta W\equiv-\,2\,g(f,\xi)\,dt_p$ with $T_0$ constant while the clock operates, then:
\begin{equation}
2\beta=\frac{1}{k}\left(\triangle\mathcal{S}+\frac{\mathcal{W}}{T_0}\right).
\end{equation}
Then, we consider that the clock produces a positive variation in entropy during its irreversible counting: $\triangle\mathcal{S}>0$, and $\mathcal{W}(x)=0$. Thus, the relation:
\begin{equation}
\triangle_{(p,p)}(\tp,p)=\int_o^\iota\left(1-e^{-\frac{1}{k}\,\triangle\mathcal{S}}\right)\,dt_p\,\geq0
\end{equation}
holds. Setting by definition $\triangle_{(p,p)}(\tp,p)\equiv\tau(\iota)-\tau(o)$, where $\tau$ is the proper time of this clock, and considering $t_p$ as the time the clock displays as an increasing counter, then at each point $x\in p$ such that $o\ll x\ll\iota$, we have the variation:
\begin{equation}
\frac{d\tau}{dt_p}=1-e^{-\frac{1}{k}\,\triangle\mathcal{S}}\,\geq0.
\label{tautp}
\end{equation}
Hence, the second law of thermodynamics would be equivalent to the monotonic increase of the proper time $\tau$ with respect to the displayed time $t_p$, with entropy linking them.
\par
The question remains as to what the verification conditions are for the chronometric hypothesis. One notes that we must start from the relation  \eqref{tautp} and the scalar potentials of metric $\tilde{g}_p(z)=r^+_p(z)\,r^-_p(z)$. The metric field $g$ is independent of $p$ and $t_p$ up to a conformal factor. In fact, the conformal factor depends upon the first-order derivatives of $t_p$ at point $z$ only such that $t_p(z)=0$. At this point $z$, one can show that the conformal factor is proportional to the square of the differential $dt_p$ and, so, of $d\tau$ from \eqref{tautp}. The mere verification of the CH would involve having a constant non-vanishing factor to set down a relation of the form $d\tau^2\equiv g$ up to a constant factor. The condition is that for every point $z\in p$ between $o$ and $\iota$, $\triangle\mathcal{S}(z)=\mathcal{S}(z)-\mathcal{S}(o)=cste$. Then, necessarily, we find $\mathcal{S}(z)=\mathcal{S}(o)$. This cannot be required for the whole of the interval between $o$ and $\iota$, and then, it remains only to suppose that there exist $o\ll z_0\ll\iota$ and $z_0\in p$ such that  $\mathcal{S}(z)=cste\,(\neq\mathcal{S}(o))$ when $z_0\ll z\ll\iota$.
Hence, only in that case, the chronometric hypothesis would be equivalent to Einstein's adiabatic hypothesis. Moreover, if $dt_p\simeq d\tau$, then it requires, first, that $\triangle\mathcal{S}(z)$ is high when $z_0\ll z\ll\iota$, and second, that we have an adiabatic process. These are conditions that atomic clocks should satisfy with $z_0$ ``closed" to $o$.
\par\smallskip
Additionally, let us investigate the case with $\mathcal{W}>0$ and $T_0$ constant. For instance, the clock starts to spin or acquire a moment of rotation. Then, to first order, we find the following expression:
\begin{equation}
\frac{d\tau}{dt_p}\simeq\frac{1}{k}\,\left(\triangle\mathcal{S}-\frac{\mathcal{W}}{T_0}\right)\,<\,
\frac{1}{k}\,\triangle\mathcal{S}.
\end{equation}
In other words, this process leads to a ``slower aging" of the clock compared to the case without rotation. This phenomenon should be strictly different from the ``rejuvenation"  case found by Prigogine \textit{et al.} \citeyearpar{prigordon2006} in the same circumstances of a rotating thermodynamic system. Nevertheless, they did not account for the time of rotation while this rejuvenating effect occurred. This time could compensate for the rejuvenation, such that the system would only age more slowly, and then, actually, their phenomenon would be very similar to the present case.
\subsection{The Hafele-Keating experiment}
This well-known experiment \citep{hafelekeat172,hafelekeat272}  brings into play three reference frames, each one using a Cesium atomic clock: two jet aircrafts, one moving westward $\tp_W$ and the other eastward $\tp_E$, and a ground control base station $p$ on the equator at null altitude. Actually, Hafele and Keating used a fourth reference frame, i.e., an inertial reference frame $q$ (thus, in particular, without rotation)  associated with a remote star. Following our notations, they computed the integrals $\triangle_{(q,q)}(\tp_W,p)$ and $\triangle_{(q,q)}(\tp_E,p)$ to the first order of the relativistic effects. The two aircrafts were supposed to fly along the equator at the same velocity $v$ with respect to the ground. The altitudes and the maximal velocities during this experiment were typically about $h\simeq10^4m$ and $v\simeq265m/s$, respectively. The Earth's equatorial radius is about $R\simeq6.38\times10^6m$ and the mean angular velocity for the Earth rotation about $\Omega\simeq7.3\times10^ {- 5}\,rad/s$. Thus, the equatorial tangential velocity is about $v_T=R\,\Omega\simeq465m/s$.
The aircraft moving towards the east moves in flight at a speed $v_E=v_T+v$ with respect to the frame $q$, and the aircraft moving towards the west travels at a speed of $v_W=v_T-v$. In addition, we denote by $\mathrm{g}$ the constant of gravity at the equator.
\par
Now, considering the first term in the integral in the formula for $\triangle_{(q,q)}(\tp_E,p)$, the term $g(\eta,\tzeta^+_1)$ in \eqref{diffet} is the factor $\gamma$ for one of the aircrafts.  For the aircraft flying toward the east, we find:
\begin{equation}
g(\eta,\tzeta^+_1)\equiv\frac{1}{\sqrt{1-\dfrac{v_E^2}{c^2}}}\,.
\end{equation}
In this expression, $g$ is the Minkowski metric, since we refer only to the remote inertial frame $q$. Then, to the first order, we obtain:
\begin{equation}
\frac{1}{g(\eta,\tzeta^+_1)}\simeq1-\frac{v_E^2}{2c^2}=1-\frac{(R\,\Omega+v)^2}{2c^2}\,.
\end{equation}
The second term in the integral can also be written to the first order as:
\begin{equation}
\frac{1}{g(\eta,\zeta^+_1)}\simeq1-\frac{v_T^2}{2c^2}=1-\frac{(R\,\Omega)^2}{2c^2}\,.
\end{equation}
Lastly, according to our interpretations, the coefficient $e^ {-\beta^-_p}$ is equal to 1, because $p$ is not subjected to an external working force. On the other hand, the coefficient $e^ {-\tilde\beta^-_ {\tp_E}}$ will reflect the work $\,\widetilde{W}$ of the external force that brings the aircraft $\tp_E$ to its cruising altitude and speed.
Under these conditions, and contrary to the function $\beta$ in the previous example for one clock only, it is necessary to take with $\tilde\beta^-_{\tp_E}\equiv\tilde\beta^-_E$ the expression $e^{-\tilde\beta^-_E}=e^{-\widetilde{W}}\simeq1-\widetilde{W}$, where $\,\widetilde{W}$ is the work per energy of mass unit $m\,c^2$, where $m$ is the mass of the aircraft.
Thus, with $\widetilde{\mathcal{W}}$ being the work of the forces (directed to the ground) given by the accelerometers and considering $\mathrm{g}$ to be roughly constant, then we find the relation: 
\begin{equation}
\widetilde{W}=\frac{\widetilde{\mathcal{W}}}{m\,c^2}\simeq-\frac{\mathrm{g}h}{c^2}.
\end{equation}
Hence, to the first order, the relation:
\begin{equation}
\frac{e^{-\tilde\beta^-_E}}{g(\eta,\tzeta^+_1)}-\frac{1}{g(\eta,\zeta^+_1)}\simeq
\frac{\mathrm{g}h}{c^2}-\frac{(2\,R\,\Omega\,v-v^2)}{2\,c^2}
\end{equation}
is exactly that given by Hafele and Keating. 
\par\smallskip
Thus, the following integrands correspond to each integral:
\begin{eqnarray}
\triangle_{(q,q)}(\tp_W,p)&\longrightarrow&\frac{\mathrm{g}h}{c^2}+\frac{(2\,R\,\Omega\,v+v^2)}{2\,c^2},\\
&&\notag\\
\triangle_{(q,q)}(\tp_E,p)&\longrightarrow&\frac{\mathrm{g}h}{c^2}-\frac{(2\,R\,\Omega\,v-v^2)}{2\,c^2}.
\end{eqnarray}
The integrals are both computed with respect to the variable $dt_q$, and thus, we can subtract them to find the differential aging $\triangle_{(q,q)}(\tp_W,p)-\triangle_{(q,q)}(\tp_E,p)$ between the airplanes. Equivalently, the term:
\begin{equation}
\frac{2\,R\,\Omega\,v}{c^2}
\label{rome}
\end{equation}
must be integrated. Rigorously, this difference is possible only if the two aircrafts are in flight at the same speed, in opposite directions, and at the same distance from the base $p$. This hypothesis is also needed to compare with the integrand in the difference $\triangle_{(q,q)}(\tp_W,q)-\triangle_{(q,q)}(\tp_E,q)$, i.e., to avoid passing by the basis $p$. Then, we easily obtain the corresponding following integrands:
\begin{eqnarray}
\triangle_{(q,q)}(\tp_W,q)&\longrightarrow&\frac{\mathrm{g}h}{c^2}+\frac{(2\,v_T\,v-v^2_T-v^2)}{2\,c^2},
\\
&&\notag\\
\triangle_{(q,q)}(\tp_E,q)&\longrightarrow&\frac{\mathrm{g}h}{c^2}-\frac{(2\,v_T\,v+v^2_T+v^2)}{2\,c^2},
\end{eqnarray}
Subtracting these two terms gives the same result \eqref{rome}.
\par
One can also consider the use of the formula \eqref{twinp}. As previously indicated, the metric $g$ is the Minkowsky metric, only far from the Earth.
In the vicinity of the Earth, this is the Schwartzschild metric, 
and from the viewpoint of each aircraft in motion or of the base station $p$, this metric is in a rotating reference frame. In addition, it must be synchronous, or equivalently, conditions such that $dt_p(\xi)=1$ must be satisfied.
\par\smallskip
The Schwarzschild metric can be approximated by the expression  ($h\ll R$):
\begin{eqnarray}
ds^2&\simeq&\left(1-\frac{r_T}{R}\left(1-\frac{h}{R}\right)\right)c^2\,dt^2
\notag\\
&&\qquad-\left(1+\frac{r_T}{R}\left(1-\frac{h}{R}\right)\right)\,dr^2
-r^2(d\theta^2+\sin(\theta)^2\,d\phi^2),
\end{eqnarray}
where $r=R+h$, and  $r_T\simeq8.87\times10^{-3}\,m$ is the gravitational radius of the Earth. In a rotating reference frame such that $\phi'=\phi-\Omega\,t'$, $t'=t$, and by replacing $r$ by $R+h$ with the approximation $h\ll R$ and taking into account the orders of magnitude for
${r_T}/{R}$, ${R\,\Omega}/{c}$
and ${h}/{R}$, we can write the following approximation of $ds$:
\begin{eqnarray}
ds^2&\simeq&\left(1-\frac{r_T}{R}\right)c^2\,dt'^2
-\left(1+\frac{r_T}{R}\right)\,dh^2\notag\\
&& \qquad-R^2\,\left(1+\frac{2\,h}{R}\right)\,\left(
2\,\Omega\,\sin(\theta)^2\,d\phi'\,dt'+d\theta^2+\sin(\theta)^2\,d\phi'^2
\right).
\end{eqnarray}
However, we need a metric in a synchronous frame, so a change of basis is performed by replacing $dt'$ by $\lambda\,dt'$ and also by multiplying all of the terms by a factor~$\mu$ (i.e., the basis is defined up to a conformal factor and a projective transformation):
\begin{eqnarray}
ds^2&\simeq&\,\mu\,\lambda^2\left(1-\frac{r_T}{R}\right)c^2\,dt'^2
-\mu\,\left(1+\frac{r_T}{R}\right)\,dh^2\notag\\
&& \qquad-\mu\,R^2\,\left(1+\frac{2\,h}{R}\right)\,\left(
2\,\lambda\,\Omega\,\sin(\theta)^2\,d\phi'\,dt'+d\theta^2+\sin(\theta)^2\,d\phi'^2
\right).
\end{eqnarray}
The constants $\mu$ and $\lambda$ are chosen in order to normalize the metric, i.e., its determinant is equal to $-1$ on the basis $(c\,dt',dh,R\,d\theta,R\,\sin(\theta)\,d\phi')$, and with the constraint:
\begin{equation}
\mu\,\lambda^2\left(1-\frac{r_T}{R}\right)=1
\end{equation}
for the value of $\mathrm{g}_{00}$. Then, to the first order,  we find that $\mu\simeq1-{4h}/{3R}$ and $\lambda\simeq1+{2h}/{3R}$, and so:
\begin{eqnarray}
ds^2&\simeq&c^2\,dt'^2
-2\,R^2\,\left(1+\frac{4h}{3R}\right)\Omega\,\sin(\theta)^2\,d\phi'\,dt'\notag\\
&& \qquad-\left(1-\frac{4h}{3R}\right)\,dh^2-R^2\left(1+\frac{2h}{3R}\right)\,(d\theta^2+\sin(\theta)^2\,d\phi'^2)\,.
\end{eqnarray}
In order to evaluate the factor $\gamma$ for this metric, the tangential velocity denoted by $v'$ of an object with respect to this rotating reference frame is such that:
\begin{equation}
v'\equiv R\,\frac{d\phi'}{dt'}\,,
\end{equation}
with the upward velocity $v'_h$ defined by:
\begin{equation}
v'_h\equiv \frac{dh}{dt'}\,.
\end{equation}
Setting $\theta=\pi/2$ at the equator, the inverse of the factor $\gamma$ for this metric is roughly of the form:
\begin{equation}
\frac{1}{\gamma}\simeq1-\frac{1}{2\,c^2}
\left\{
v'^2_h+2\,v_T\,v'+v'^2
+\frac{2h}{3R}
\left[
v'^2+4\,v_T\,v'-2v'^2_h
\right]
\right\}.
\end{equation}
Let us now consider the formula \eqref{diffet} to compute $\triangle_{(p,p)}(\tp_W,\tp_E)$. We assume the angles are $\phi'>0$ toward the east.
With $v'=v_E=v$ or $v'=v_W=-v$ and assuming $v'_h=0$, then, for the two cases, we find the relations:
\begin{eqnarray}
\tilde\beta^-_W=-\frac{\mathrm{g}\,h}{c^2},
&&
\frac{e^{-\tilde\beta^-_W}}{g(\eta,\tzeta^+_1)}\simeq1+\frac{1}{c^2}
\left\{
\mathrm{g}\,h+v\,v_T-\frac{v^2}{2}
+\frac{h}{3R}
\left[
4\,v\,v_T-v^2
\right]
\right\},\\
&&\notag\\
\beta^-_E=-\frac{\mathrm{g}\,h}{c^2},
&&
\frac{e^{-\beta^-_E}}{g(\eta,\zeta^+_1)}\simeq
1+\frac{1}{c^2}
\left\{
\mathrm{g}\,h-v\,v_T-\frac{v^2}{2}
-\frac{h}{3R}
\left[
4\,v\,v_T+v^2
\right]
\right\}.
\end{eqnarray}
Hence, in order to obtain $\triangle_{(p,p)}(\tp_W,\tp_E)$, we need to integrate:
\begin{equation}
\frac{2}{c^2}\left(1+\frac{4h}{3R}\right)\,v\,v_T
\end{equation}
with respect to $t_p$. At the zeroth order in $h$, we find again the formula
\eqref{rome}.
\par\bigskip
We consider now the formulas obtained from the viewpoint of each aircraft, that is to say, $\triangle_{(\tp_W,\tp_W)}(\tp_W,\tp_E)$ or $\triangle_{(\tp_E,\tp_E)}(\tp_W,\tp_E)$. First of all, in their proper frames, none of the external forces work, since the origins of the reference frames are the aircrafts themselves. However, the functions $\beta$ are the evaluations of the work of the forces being applied on the aircraft which is not at the origin of the considered reference frame. Formula \eqref{twinp} must be applied:
\begin{equation}
\triangle_{(\tp_W,\tp_W)}(\tp_W,\tp_E)=\int_o^\iota \left\{
e^{-\beta_W}-
\frac{e^{-\tilde\beta^-_{E}}}{g(\xi^+_1,\xi)}
\right\}\,dt_{\tp_W}\,.
\end{equation}
There are the following correspondences: $\beta_W$ is the work of the forces given by the accelerometers placed on the aircraft moving toward the west and estimated in the reference frame of the same aircraft. Reciprocally, $\tilde\beta^-_{E}$ is the work of the forces given by the accelerometers placed on the aircraft moving to the east and estimated by the plane flying toward the west. From the point of view of each airplane, these two sources of work would be the same if they could simultaneously be estimated with the function $F$. However, in fact, this is not even necessary because the function $\tilde\beta^-_{W}$ takes into account the temporal delay via the map $f^-_{\tp_W}$.
\par\medskip
For the two differences, it is thus necessary to estimate the coefficients
$g(\xi^+_1,\xi)$ ascribed to  $\tp_E$. There is the following correspondence:
\begin{equation}
\frac{1}{g(\xi^+_1,\xi)}\simeq1-\frac{1}{2\,c^2}
\left\{
{v'^2_h}+2\,v_T\,{v'_E}+{v'^2_E}
+\frac{2h}{3R}
\left[
{v'^2_E}+4\,v_T\,{v'_E}-2{v'^2_h}
\right]
\right\},
\end{equation}
where $v'_E$, a function of $t_{\tp_W}$, is the velocity of the airplane travelling towards the east from the point of view of $\tp_W$, and $v'_h$ is its upward velocity. The expressions for the functions $\beta$ are:
\begin{equation}
\tilde\beta^-_{E}=-\frac{1}{c^2}\,\mathrm{g}\,h_E(t_{\tp_W}),
\qquad
\beta_W=-\frac{1}{c^2}\,\mathrm{g}\,h_W(t_{\tp_W}),
\end{equation}
where the altitudes are functions of the values of the clock function 
$t_{\tp_W}$, $h_E$ is the altitude of the airplane moving toward the east, and $h_W$ is that for the plane traveling west. These altitudes are, actually, relative altitudes between the two airplanes, not respective to the ground.
One can thus make the reasonable assumption that $h_W\simeq h_E\simeq 0$. Under these conditions, by considering $v'_E=2\,v$, the integrand of $\triangle_{(\tp_W,\tp_W)}(\tp_W,\tp_E)$ is of the approximate form:
\begin{equation}
\frac{2}{c^2}\left\{
v^2+v\,v_T+\frac{2h}{3R}\left[v^2+2\,v\,v_T\right]
\right\}.
\end{equation}
This expression is far from that obtained in the reference frames of $p$ or $q$, but in fact it means that the comparison of the integrated terms is no longer valid. It is the complete integral that must be computed.
It follows that the Doppler terms in $v^2$ must include dilatations and contractions of the durations depending on whether the airplanes are moving away from each other or are approaching one another along the equator. The contributions of these two modifications in duration will be the same and will be cancelled out at the conclusion of a circumnavigation. Thus, only the effect of the term in $v\,v_T$ will persist, leading to the same final result.
\section{Conclusion}
We have seen  that the clock paradox is perfectly solved within the relativity framework due to the asymmetry resulting from the factors 
$e^\beta$. However, at the same time, it is necessary to ascribe a new meaning to these conformal factors. These factors are strongly associated with the concept of total energy minus kinetic energy, which is not necessarily defined by a potential along each timelike worldline (particles).
Actually, these functions $\beta$ imply in themselves the data of an additional dimension.
More precisely, we have sheafs of germs of functions $\beta$ that are classes of scalar fields, which ``mimic", at each given event, the values of the functions $\beta$. In other words, we could have no such scalar fields on $\m$, but there is always the need for non-vanishing conformal scale factors, i.e., values independent on the \textit{locus} in the spacetime, which is an extra dimension. In the same vein, we can suggest a reply to a concluding question of Ehlers \textit{et al.}, asking whether other interpretations of the so-called Weyl \textit{streckenkr\"ummung} bivector $\mathbf{F}$ rather than those ascribing the latter to the electromagnetic field \textit{``might contain some physical truth"}. The fact is that as a field, per se, we may have no such field $\mathbf{F}$, but only germs of such fields along particles or beams of particles. Again, this bivector $\mathbf{F}$ would be related as a germ to derivatives of the functions $\beta$. However, in any case, such (bi)vectors would be defined only in jet manifolds of germs of functions and, as such, would be associated with an extra variable.
\par
Thus, it seems necessary to consider a spacetime supplied with an extra dimension of energy, i.e., the conformal scale factors, that translates the various physical transformations without gravitational origins. Therefore, it is basically a ``spacetime" of five dimensions upon a conformal spacetime of four dimensions. However, the geometrical structure should also be a product of copies of the spacetime $\m$ fibered by conformal scale factors (or equivalently, $F$).
In conclusion, this extra dimension is absolutely necessary in order to provide an account of the internal evolution of objects in relation to the spacetime structure, and it alone gives a meaning to the concept of the proper time of an object in relation to the chronology given on a spacetime.


\begin{thebibliography}{17}
\expandafter\ifx\csname natexlab\endcsname\relax\def\natexlab#1{#1}\fi
\expandafter\ifx\csname url\endcsname\relax
  \def\url#1{\texttt{#1}}\fi
\expandafter\ifx\csname urlprefix\endcsname\relax\def\urlprefix{URL }\fi

\bibitem[{Allan(1966)}]{allan66}
Allan, D.~W. (1966). Statistics of atomic frequency standards. 
\textit{Proceedings of the IEEE}, 54, 221--230.

\bibitem[{Audretsch(1983)}]{audret83}
Audretsch, J. (1983). Riemannian structure of space-time as a
consequence of quantum mechanics. \textit{Physical Review D}, 27, 2872--2884.

\bibitem[{Ehlers et~al.(1972)Ehlers, Pirani, and Schild}]{ehl72}
Ehlers, J., Pirani, F.~A.~E., \& Schild, A. (1972). The geometry of free fall and light propagation. In L. O{'}Raifeartaigh (Ed.), \textit{General relativity, papers in honour of J.~L. Synge} (pp. 63--84). Oxford: Clarendon Press.

\bibitem[{Einstein(1918)}]{Einstein18}
Einstein, A. (1918). Dialog {\"u}ber einw{\"a}nde gegen die
relativit{\"a}tstheorie. \textit{Die Naturwissenschaften}, 6, 697--702.

\bibitem[{Ghins and Budden(2001)}]{ghinsbuden2001}
Ghins, M., \& Budden, T. (2001). The equivalence principle. \textit{Studies in History and Philosophy of Modern Physics}, 32, 33--51.

\bibitem[{Hafele and Keating(1972{\natexlab{a}})}]{hafelekeat172}
Hafele, J.~C., \& Keating, R.~E. (1972{\natexlab{a}}). Around-the-world atomic clocks: Predicted relativistic time gains. \textit{Science}, 177, 166--168.

\bibitem[{Hafele and Keating(1972{\natexlab{b}})}]{hafelekeat272}
Hafele, J.~C., \& Keating, R.~E. (1972{\natexlab{b}}). Around-the-world atomic clocks: Observed relativistic time gains. \textit{Science}, 177, 168--170.

\bibitem[{Konheimer and Penrose(1967)}]{kronpen67}
Konheimer, E.~H., \& Penrose, R. (1967). On the structure of causal spaces.
\textit{Proceedings of the Cambridge Philosophical Society}, 63, 481--501.

\bibitem[{Kundt and Hoffmann(1962)}]{kundhoff62}
Kundt, W., \& Hoffmann, B. (1962). Determination of gravitational standard time. In \textit{Recent Developments in General Relativity - A book dedicated to Leopold Infeld's 60th birthday} (pp. 303--306). New York: Pergamon Press.

\bibitem[{Malament(1977{\natexlab{a}})}]{malam77bis}
Malament, D. (1977{\natexlab{a}}). Causal theories of time and the
conventionality of simultaneity. \textit{No{\^u}s}, 7, 293--300.

\bibitem[{Malament(1977{\natexlab{b}})}]{malam77}
Malament, D. (1977{\natexlab{b}}). The class of continuous timelike curves
determines the topology of spacetime. \textit{Journal of Mathematical Physics}, 18, 1399--1404.

\bibitem[{Misner et~al.(1973)Misner, Thorne, and Wheeler}]{misnerwheel73}
Misner, C.~W., Thorne, K.~S., \& Wheeler, J.~A. (1973). \textit{Gravitation}. San Francisco: W. H. Freeman and Compagny.

\bibitem[{Prigogine and Ordonez(2006)}]{prigordon2006}
Prigogine, I., \& Ordonez, G. (2006). Acceleration and entropy: A macroscopic
  analogue of the twin paradox. In F.~F. Orsucci and N. Sala (Eds.), \textit{New Research on Chaos and Complexity} (pp. 5--20). New York: Nova Science Publishers Inc..

\bibitem[{Unnikrishnan(2005)}]{unnik2005}
Unnikrishnan, C.~S. (2005). On {E}instein{'}s resolution of the twin clock
paradox. \textit{Current Science}, 89, 2009--2015.

\bibitem[{Wheller(1990)}]{wheeler90}
Wheller, J.~T. (1990). Quantum measurement and geometry. \textit{Physical Review D}, 41, 431--441.

\bibitem[{Woodhouse(1973)}]{wood73}
Woodhouse, N. M.~J. (1973). The differential and causal structures of
space-time. \textit{Journal of Mathematical Physics}, 14, 495--501.

\bibitem[{Zeghib(2004)}]{zeghib2004}
Zeghib, A. (2004). Lipschitz {R}egularity in {S}ome {G}eometric {P}roblems.
\textit{ Geometriae Dedicata}, 107, 57--83.

\end{thebibliography}
\end{document}